\documentclass[12pt,journal,draftclsnofoot,onecolumn]{IEEEtran}
\makeatletter
\def\ps@headings{%
\def\@oddhead{\mbox{}\scriptsize\rightmark \hfil \thepage}%
\def\@evenhead{\scriptsize\thepage \hfil \leftmark\mbox{}}%
\def\@oddfoot{}%
\def\@evenfoot{}}
\makeatother
\usepackage{multirow}
\usepackage{amssymb}
\usepackage{amsmath}
\usepackage{amsfonts}
\usepackage{graphicx}
\usepackage{subfigure}
\usepackage{cite}
\usepackage{enumerate}
\usepackage{url}
\usepackage{bm}
\usepackage{clrscode}
\usepackage{subfigure}
\usepackage{amsthm}
\usepackage[ruled]{algorithm2e}
\usepackage{algorithmic}
\allowdisplaybreaks[4]
\begin{document}
\title{Beyond Intelligent Reflecting Surfaces: \\Reflective-Transmissive Metasurface \\Aided Communications for \\Full-dimensional Coverage Extension}
\author{\IEEEauthorblockN{
		{Shuhang Zhang}, \IEEEmembership{Student Member, IEEE},
		{Hongliang Zhang}, \IEEEmembership{Member, IEEE},
		{Boya Di}, \IEEEmembership{Member, IEEE},
        {Yunhua Tan},
        {Zhu Han}, \IEEEmembership{Fellow, IEEE},
        and {Lingyang Song}, \IEEEmembership{Fellow, IEEE}
        }\\
\thanks{S. Zhang, Y. Tan, and L. Song are with Department of Electronics, Peking University, Beijing 100871 China (email: \{shuhangzhang, tanggeric, lingyang.song\}@pku.edu.cn).}
\thanks{H. Zhang is with Department of Electronics, Peking University, Beijing 100871 China, and also with Department of Electrical Engineering, Princeton University, Princeton, NJ 08544 USA (email: hongliang.zhang92@gmail.com).}
\thanks{B. Di is with Department of Electronics, Peking University, Beijing 100871 China, and also with Department of Computing, Imperial College London, London SW7 2BU, U.K. (email: diboya92@gmail.com).}
\thanks{Z. Han is with Electrical and Computer Engineering Department, University of Houston, TX 77004, USA, and also with the Department of Computer Science and Engineering, Kyung Hee University, Seoul 02447 South Korea (email: zhan2@uh.edu).}}

%\vspace{-.5cm}}

%\thanks{M. D. Renzo is with the Laboratoire des Signaux et Systemes, CNRS, CentraleSupelec, Univ Paris-Sud, Universite Paris-Saclay, 91192 Gif-surYvette, France (e-mail: marco.direnzo@l2s.centralesupelec.fr).}
%\thanks{H.~V.~Poor is with the Department of Electrical Engineering, Princeton University, Princeton, NJ 08544 USA. E-mail: poor@princeton.edu.}

\maketitle

\begin{abstract}
In this paper, we study an intelligent omni-surface~(IOS)-assisted downlink communication system, where the link quality of a mobile user~(MU) can be improved with a proper IOS phase shift design. Unlike the intelligent reflecting surface~(IRS) in most existing works that only forwards the signals in a reflective way, the IOS is capable to forward the received signals to the MU in either a reflective or a transmissive manner, thereby enhancing the wireless coverage. We formulate an IOS phase shift optimization problem to maximize the downlink spectral efficiency~(SE) of the MU. The optimal phase shift of the IOS is analysed, and a branch-and-bound based algorithm is proposed to design the IOS phase shift in a finite set. Simulation results show that the IOS-assisted system can extend the coverage significantly when compared to the IRS-assisted system with only reflective signals.
\end{abstract}

\begin{IEEEkeywords}
Intelligent omni-surface, IOS phase shift design, coverage extension.
\end{IEEEkeywords}
%\vspace{-.5cm}
\section{Introduction}
With the development of meta-surfaces, the intelligent reflecting surface~(IRS) is considered as a promising technique for future communications, since it is cost-effective to achieve a high spectral and energy efficiency~\cite{R2019}. The IRS contains a large number of elements with controllable electromagnetic responses that can shape the propagation environment into a desirable form~\cite{EZSSHL2020}, thus enhancing the quality of the communication links~\cite{HZDLSLHP2019}. In the literature, some initial works have studied the utilization of wave-reflective IRS to assist the wireless communication networks. In~\cite{HZADY2019}, a joint power allocation and continuous phase shift design has been studied in a reflective IRS-assisted system to maximize energy efficiency. In~\cite{ZDSH2020}, the achievable data rate of a reflective IRS-assisted communication system has been evaluated, and the effect of limited phase shifts on the data rate has been investigated. However, in the existed works, the signal arrived at the IRS is considered to be reflected completely. As a result, the receivers on the other side of the IRS are shielded, which leads to an incomplete wireless coverage.

In this paper, we study an intelligent omni-surface~(IOS), whose received signals can be transmitted and reflected to both sides concurrently. We utilize the IOS in a communication system to extend the service coverage for downlink transmissions. The IOS provides an ubiquitous wireless coverage for the mobile user~(MU) on its either side, and the propagation environment of the MU can be adjusted via the phase shifts of the electrically controllable elements on the IOS.

The implementation of the IOS-assisted communication system has also brought some new challenges. \emph{First}, the power of reflective and transmissive signals of the IOS may not be symmetric, i.e., the channel model of the reflective signal and the transmissive signal can be different. Therefore, the studies on the reflective IRS-assisted communications cannot be applied directly in the IOS-assisted ones. \emph{Second}, the direct communication link from the base station~(BS) to the MU may exist concurrently with the reflective and transmissive links in the IOS-assisted communication system. Such a superposition impact of multiple communication links should be considered for IOS phase shift design.

To deal with the above challenges, we first propose a novel model for the IOS with a controllable electromagnetic response on both sides, and then introduce its physical characteristics. Based on the channel model of the proposed IOS, we formulate a downlink communication spectral efficiency~(SE) maximization problem, where an IOS is utilized to improve the SE of a MU by proper phase shift design. The optimal solution to the IOS phase shift is analysed, and a branch-and-bound algorithm is proposed to design the IOS phase shifts in a finite set. The performances in terms of coverage and SE improvement are provided by simulations.

The rest of this paper is organized as follows. In Section~\ref{IOS sec}, we introduce the IOS. In Section~\ref{System Model Sec}, we present an IOS-assisted downlink communication system, including the channel model and the SE, and formulate a SE maximization problem with IOS phase shift design. An IOS phase shift design algorithm is provided in Section~\ref{Algorithm Sec}. Numerical results in Section~\ref{Simulation Sec} evaluate the performance of the proposed algorithm. Finally, we draw the conclusions in Section~\ref{Conclusion Sec}.
%\vspace{-.3cm}
\section{Intelligent Omni-Surface}\label{IOS sec}
\begin{figure}[t]
\centering
\includegraphics[width=2.6in]{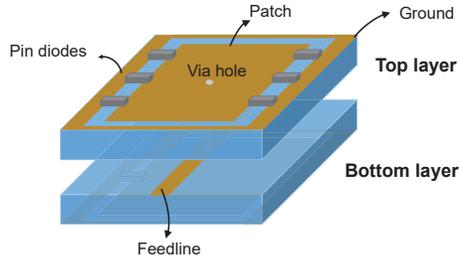}\vspace{-.3cm}
\caption{Schematic structure of an IOS element.}\vspace{-.3cm}
\label{element}
\end{figure}
The IOS is a two-dimensional array of electrically controllable IOS elements. Each element in the IOS is of the same size, with $\delta_x$ being the width and $\delta_y$ being the height, and is composed of multiple metal patches and $N_D$ PIN diodes assembled on dielectric substrates. As shown in Fig.~\ref{element}, the metal patches are connected to the ground via PIN diodes, and can be switched between \textit{ON} and \textit{OFF} states according to the applied bias voltages, based on which a unique phase shift can be added to the transmissive and reflective signals. There are $2^{N_D}$ possible phase shifts in total for each element. For generality, we assume that a subset of possible phase shifts are available, which is referred to as the \emph{available phase shifts set}, denoted by $\mathcal{S}_a=\{1,\cdots,S_a\}$. We denote the phase shifts of the $m$th element by $s_m \in \mathcal{S}_a$. We can control the PIN diodes to generate $S_a$ patterns of phase shifts for each element, with a uniform interval $\Delta \psi_m=\frac{2\pi}{S_a}$~\cite{ZDSH2020}. The possible phase shift value can be given as $l_m \Delta \psi_m$, where $l_m$ is an integer satisfying $0\leq l_m\leq S_a-1$. The phase shift of the IOS is defined as the vector of the phase shifts of all IOS elements, i.e., $\bm{s}=(s_1,\dots,s_M)$, where $M$ denotes the number of IOS elements. When a signal arrives at an IOS element from either side of it, a part of the signal transmits through the elements as the \emph{transmissive signal}, and the other part of the signal is reflected as the \emph{reflective signal}~\cite{AB2019}. The phase shift of the IOS determines the waveform of the transmissive and reflective signals concurrently.

\begin{figure}[t]
\centering
\includegraphics[width=2.6in]{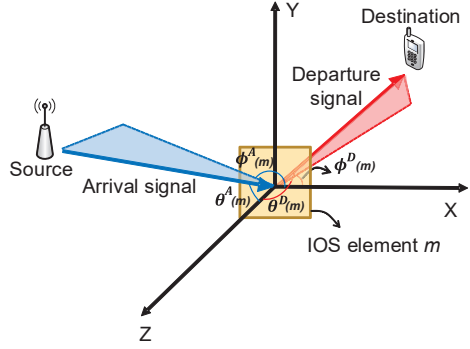}\vspace{-5mm}
\caption{Illustration for the angles of arrival and departure signals for a IOS element.}\vspace{-5mm}
\label{angle}
\end{figure}

As illustrated in Fig.~\ref{angle}, we denote the direction of the arrival signal from the source node to element $m$, and the direction from element $m$ to the destination node by $\xi^{A}(m)=(\theta^{A}(m), \phi^{A}(m))$ and $\xi^{D}(m)=(\theta^{D}(m), \phi^{D}(m))$, respectively. The influence of element $m$ on the arrival signal is denoted by a complex number $g_m$, which we refer to as the \emph{power gain} of the signal. The value of $g_m$ for a transmission link is affected by the direction of the arrival signal from the source node, i.e., $\xi^{A}(m)$, the direction of the departure signal to the destination node, i.e., $\xi^{D}(m)$, and the element phase shifts $s_m$. The experimental equation of the corresponding power gain can be expressed as~\cite{TCCDHRZJCC2019}
\vspace{-.2cm}
\begin{equation}\label{signal gain}\vspace{-.2cm}
\begin{split}
g_m(&\xi^A(m),\xi^D(m),s_m)=\\
&\sqrt{G_m K^{A}(m) K^{D}(m)\delta_x\delta_y|\gamma_m|^2}\exp{(-j \psi_{m})},
\end{split}
\end{equation}
where $G_m$ is the antenna gain of element $m$, and $\psi_{m}$ is the corresponding phase shift. Variable $\gamma_m$ is the power ratio of the departure signal to the arrival signal. It can either be a function of $s_m$ or a constant, which is related to the schematic structure of the IOS element. $K^{A}(m)$ and $K^{D}(m)$ are the normalized power radiation pattern of the arrival signal and departure signal, respectively.\footnote{The departure signal can be either transmissive signal or reflective signal.} An example of the normalized power radiation pattern is given as following:
\vspace{-.2cm}
\begin{align}\label{Ki}\vspace{-.2cm}
K^{A}(m)=|\cos^3{\theta^{A}(m)}|,
\end{align}
\vspace{-.2cm}
\begin{align}\label{Ko}\vspace{-.2cm}
	K^{D}(m)=\left\{
	\begin{aligned}
		&|\cos^3{\theta^{D}(m)}|,~\theta^{D}(m)\in(0,\pi/2),\\
		&\epsilon|\cos^3{(\pi-\theta^{D}(m))}|,~\theta^{D}(m)\in(\pi/2,\pi),
	\end{aligned}
    \right.
\end{align}
where $\epsilon$ is a constant parameter that describes the power ratio of the transmissive signal to the reflective signal\footnote{The value of $\epsilon$ is determined by the material of the IOS element~\cite{PG2013}, which needs to be performed before the IOS deployment. Therefore, the distribution of the MU should be estimated before the IOS deployment.}. %It is worthwhile to mention that the strength of both the reflective and the transmissive signals satisfies~(\ref{Ko}), where $\theta^{D}(m)\in(0,\pi/2)$ shows the power gain of the reflective signal, and $\theta^{D}(m)\in(\pi/2,\pi)$ shows the power gain of the transmissive signal.

\vspace{-.3cm}
\section{System Model and Problem Formulation}\label{System Model Sec}
In this section, we first describe an IOS assisted single MU downlink system, and then introduce the channel model and SE of the IOS-assisted system. Finally, we formulate a downlink SE maximization problem by optimizing the IOS phase shift.
\vspace{-.3cm}
\subsection{Scenario Description}
\begin{figure}[!tpb]
\centering
\includegraphics[width=3in]{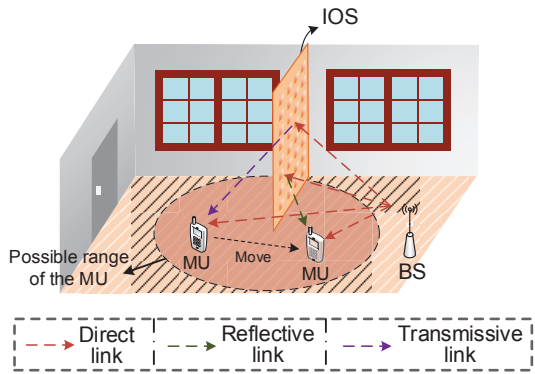}%\vspace{-8mm}
\caption{System model for the IOS-aided downlink cellular system.}%\vspace{-3mm}
\label{model}
\end{figure}
As shown in Fig.~\ref{model}, we consider a downlink transmission scenario in an indoor environment, which consists of one BS and one MU. We assume that the MU is randomly located in a range denoted by $\mathcal{L}$. Due to the severe channel fading and complicated scattering in the indoor environment, the MU may suffer low quality of service of the communication link to the BS. To tackle this problem, we deploy an IOS in the indoor environment. The power of the received signals at the MU can be improved with either the transmissive signal or the reflective signal of the IOS, and the type of the received signal from the IOS is determined by the current location of the MU. The IOS can be viewed as an antenna array far away from the BS, inherently capable of realizing beamforming via the IOS phase shift design, which will be introduced in detail in Section~\ref{beamforming subsection}.
\vspace{-.2cm}
\subsection{Channel Model}\label{interference sec}
The channel from the BS to the MU consists of two parts: the direct path from the BS to the MU bypassing the IOS, and the reflective-transmissive channel that goes through the IOS. The detailed descriptions of the direct path and the reflective-transmissive channel are given in the following.

\subsubsection{Direct Path Bypassing the IOS}
The channel model of the direct path from the BS to the MU is similar to the one in conventional cellular networks, which can be formulated as a Rician channel. The channel of the direct path from the BS to the MU can be written by
\vspace{-.2cm}
\begin{align}\label{direct path equation}\vspace{-.2cm}
h_{D}=\sqrt{\frac{\kappa}{1+\kappa}}h_D^{LoS}+\sqrt{\frac{1}{1+\kappa}}h_D^{NLoS},
\end{align}
where $\kappa$ is the Rician factor indicating the ratio of the LoS component to the non-line-of-sight (NLoS) one, and $h_D^{LoS}$ and $h_D^{NLoS}$ are the LoS and NLoS components of the direct path, respectively. According to~\cite{G2005}, the LoS component of the normal channel between the BS and the MU can be given as $h_D^{LoS}=\sqrt{G^{tx}G^{rx}d_{BS,MU}^{-\alpha}}\exp{(-j \frac{2\pi}{\lambda}d_{BS,MU})}$, where $G^{tx}$ is the transmission antenna gain of the BS antenna, $G^{rx}$ is the receiving antenna gain of the MU, $d_{BS,MU}$ is the distance from the BS to the MU, and $\alpha$ is the path-loss parameter. Similarly, the NLoS component can be written as $h_D^{NLoS}=PL(d_{BS,MU})h^{SS}$, where $PL(\cdot)$ is the channel gain for the NLoS component, and $h^{SS}\sim\mathcal{CN}(0,1)$ denotes the small-scale NLoS components.

\subsubsection{Transmissive-Reflective Channel via the IOS}
As introduced in Section~\ref{IOS sec}, the departure signal of the IOS contains two parts: transmissive signal and reflective signal. The location of the MU determines whether it receives the transmissive signal or the reflective signal. The channel from the BS to the MU via the IOS can be considered as the sum of $M$ channels from the BS to the MU via every IOS element. Since the BS-IOS-MU link is much stronger than the NLoS ones, the channel from the BS to the MU via each IOS element can be modeled as a Racian channel. The channel gain from the BS to the MU via IOS element~$m$ is given as
\vspace{-.2cm}
\begin{align}\label{BS-IOS-MU channel}\vspace{-.2cm}
h_m=\sqrt{\frac{\kappa}{1+\kappa}}h_m^{LoS}+\sqrt{\frac{1}{1+\kappa}}h_m^{NLoS}.
\end{align}
The LoS component of $h_m$ can be expressed as
\vspace{-.1cm}
\begin{equation}\label{main_channel}\vspace{-.1cm}
\begin{split}
h_m^{LoS}=&\frac{\lambda \sqrt{G^{tx} F_{m}^{tx} G^{rx} F^{rx}} \exp\Big(\frac{-j2\pi(d_{BS,m}+d_{m,MU})}{\lambda}\Big)}{(4\pi)^{\frac{3}{2}}d_{BS,m}d_{m,MU}}\\ \times& g_m(\xi^{A}(m),\xi^{D}(m),s_m),
\end{split}
\end{equation}
where $\lambda$ is the wave length corresponding to the carrier frequency, $G^{tx}$ and $G^{rx}$ are antenna gains of the BS and the MU, respectively, $F_{m}^{tx}$ is the normalized power gain of the BS antenna in the direction of the $m$-th IOS element, $F_{m}^{rx}$ is the normalized power gain of the MU in the direction of IOS element~$m$, $d_{BS,m}$ and $d_{m,MU}$ are the distances from the IOS element~$m$ to the BS and to the MU, respectively. $g_m(\xi^{A}_k(m),\xi^{D}(m),s_m)$ is the power gain of the signal toward the MU via IOS element~$m$, which is given in~(\ref{signal gain}). The NLoS component of $h_m$ can be written as
\vspace{-.1cm}
\begin{equation}\vspace{-.1cm}
h_m^{NLoS}=PL(d_{BS,m})PL(d_{m,MU})h^{SS},
\end{equation}
where $PL(\cdot)$ is the channel gain for the NLoS component, and $h^{SS}\sim\mathcal{CN}(0,1)$ denotes the small-scale NLoS component.

In summary, the channel gain from the BS to the MU can be written as
\vspace{-.3cm}
\begin{align}\label{channel equation}\vspace{-.6cm}
h=\sum_{m=1}^M h_m+h_{D},
\end{align}
where the first term represents the superposition of the transmissive-reflective channel of the $M$ IOS elements, and the second term is the direct path.
\vspace{-.5cm}
\subsection{Spectral Efficiency with IOS Phase Shift Design}\label{beamforming subsection}\vspace{-.1cm}
In this part, we introduce the IOS phase shift design, with which the power of the signal received by the MU can be improved significantly. According to~(\ref{channel equation}), the received signal at the MU can be expressed as
\vspace{-.2cm}
\begin{equation}\vspace{-.1cm}
z=\sum_{m=1}^M h_m x+h_{D} x+n,
\end{equation}
where $n$ is the additive white Gaussian noise~(AWGN) at the MU with zero mean and $\sigma^2$ as the variance, and $x$ denotes the transmitted signal with $|x|^2=1$.

The downlink SE of the MU can be defined as the data rate in bits per second per Hz, which is given as
\vspace{-.1cm}
\begin{equation}\label{rate}\vspace{-.1cm}
R=\log_2{\left(1+\frac{P|h|^2}{\sigma^2}\right)},
\end{equation}
where $P$ is the transmission power of the BS, which is given as a constant in this paper.
\vspace{-.3cm}
\subsection{Problem Formulation}\label{Problem Formulation Sec}
In this part, we formulate an IOS phase shift design problem to maximize the average downlink SE of the system. As shown in~(\ref{main_channel}) and~(\ref{rate}), the SE of the MU is determined by its location and the phase shift of the IOS $\bm{s}$. Given the MU location $\bm{l}$, the signal-to-noise ratio at the MUs can be added instructively with proper IOS phase shift design, and the SE can be improved. In the following, we aim to maximize the SE of the MU randomly distributed in the range of $\mathcal{L}$ by optimizing the phase shift of IOS $\bm{s}$, and then the problem can be formulated as
\begin{subequations}\label{formulate problem}
\begin{align}
\max_{\bm{s}} \ & R, \forall \ \bm{l} \in \mathcal{L},\\
\label{cons1}
s.t. \ & s_m \in \mathcal{S}_a, m=1, 2, \cdots M.
\end{align}
\end{subequations}
Constraint (\ref{cons1}) is the feasible set for the phase shifts of each IOS element.
\vspace{-.3cm}
\section{IOS Phase Shift Design}\label{Algorithm Sec}
In this section, we first relax the variable in problem~(\ref{formulate problem}) and discuss the SE maximization problem with continuous IOS phase shifts, and then propose a branch-and-bound based algorithm to design the IOS phase shift in finite set $\mathcal{S}_a$.
\vspace{-.3cm}
\subsection{Continuous IOS Phase Shift Design}
\begin{figure*}[t]
    \centering
    \includegraphics[width=4.5in]{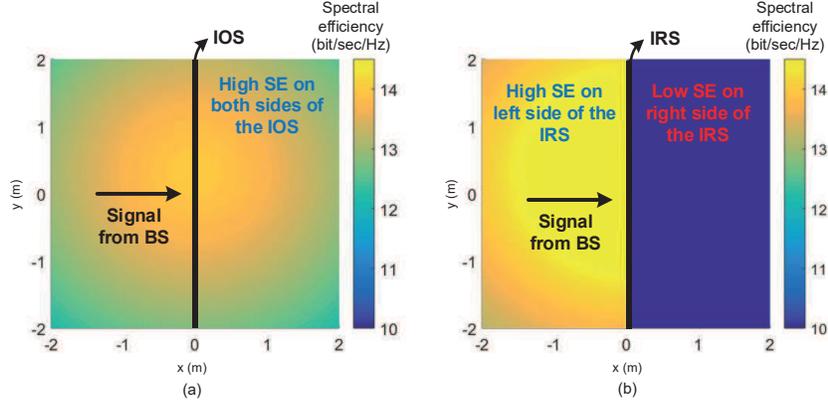}\vspace{-4mm}
    \caption{Simulation for maximum SE. (a) IOS system. (b) IRS system.}\vspace{-6mm}
    \label{simulation0}
\end{figure*}
As shown in~(\ref{direct path equation}), the channel of the direct path is not affected by the phase shift of the IOS, and thus we only need to optimize the SE of the BS-IOS-MU link. Moreover, the BS-IOS-MU link is modeled as a Racian channel given in~(\ref{BS-IOS-MU channel}), whose NLoS component is not affected by the phase shift of the IOS. Therefore, problem~(\ref{formulate problem}) can be simplified as maximizing the SE of the LoS component of the BS-IOS-MU link, which is shown as
%\vspace{-.3cm}
\begin{subequations}\label{simplified problem}\vspace{-.3cm}
\begin{align}
\max_{\bm{s}} \ &\log_2{\left(1+\frac{P|\sum_{m=1}^M h_m^{LoS}+h_D|^2}{\sigma^2}\right)}, \forall \ \bm{l} \in \mathcal{L},\\
s.t. \ & s_m \in \mathcal{S}_a, m=1, 2, \cdots M.
\end{align}
\end{subequations}

Problem~(\ref{simplified problem}) is an integer optimization problem, which cannot be solved by the optimization methods for continuous variables. In the following, we first solve the problem with $\bm{s}$ being relaxed to a continuous variable, and then we propose a branch-and-bound based algorithm to design the discrete phase shift for each IOS element.

\textbf{Proposition 1:} When the power ratio of the departure signal to the arrival signal $\gamma_m$ is considered as a constant, the optimal phase shift of each element satisfies\vspace{-.1cm}
\begin{equation}\vspace{-.1cm}
\psi_m=\frac{2\pi}{\lambda}(d_{BS,MU}-d_{BS,m}-d_{m,MU}), m=1, \cdots, M.
\end{equation}
\begin{proof}
See Appendix A.
\end{proof}
\vspace{-.4cm}
\subsection{Finite IOS Phase Shift Design}\vspace{-.1cm}
The optimal solution proposed in Proposition 1 cannot be obtained by the IOS elements with a finite set of phase shift $\mathcal{S}_a$. For IOS element $m$, we denote its optimal phase shift by $\psi_m^{opt}$. For IOS element~$m$, its phase solution to problem~(\ref{formulate problem}) is one of the two consecutive phase shifts of $s_m$ and $s_{m+1}$, which satisfies $\psi_m\leq\psi_m^{opt}\leq \psi_{m+1}$. In the following, we propose a branch-and-bound based algorithm to solve the IOS phase shift design in the finite set $\mathcal{S}_a$ efficiently.

The solution space of IOS phase shift~$\bm{s}$ can be considered as a binary tree structure. Each node of the tree contains the phase shift of all the IOS elements, i.e., $\bm{s}=(s_1,\dots,s_M)$. At the root node, all the variables in~$\bm{s}$ are unfixed. The value of an unfixed variable at a father node can be either $s_m$ or $s_{m+1}$, which branches the node into two child nodes. The objective of the proposed algorithm is to search the tree for the optimal solution of problem~(\ref{formulate problem}) with the following three steps.

\textbf{Step 1: Initialization.} We first set a random IOS phase shift, and the corresponding SE is given as the lower bound of the solution.

\textbf{Step 2: Bound Calculation.} We then start to search the optimal solution from the root node. On each node we first evaluate the upper bound of the objective function with variable relaxation, and the upper bound can be calculated with the phase shift solution proposed in Proposition 1.

\textbf{Step 3: Variable Fixation.} In this step, we prune the branches whose upper bounds are below the value of the current solution. When a solution that outperforms the current solution is found, we replace the current solution with the new one, and continue the branch-and-bound algorithm. The phase shifts of an element is fixed when only one feasible value satisfies the bound requirements.

The algorithm terminates when all the element phase shifts are fixed, and the corresponding current solution is the final solution of problem~(\ref{formulate problem}). The proposed algorithm that solves problem~(\ref{formulate problem}) is summarized as Algorithm~\ref{IOS Configuration}.

\begin{algorithm}[!t]
\caption{IOS Phase Shift Design Algorithm.}
\begin{algorithmic}[1]\label{IOS Configuration}
\STATE {\textbf{Initialization:} Compute an initial solution $\bm{s}$ to problem~(\ref{simplified problem}) and set the SE as the lower bound $R_{lb}$}
\STATE {\textbf{While} Not all nodes are visited or pruned}
\STATE {\quad Calculate the upper bound of the current node~$R_{ub}$}
\STATE {\quad \textbf{If} $R_{ub}<R_{lb}$: Prune this branch}
\STATE {\quad \textbf{If} Variable $s_m$ is fixed: Go to the node with $s_m$}
\STATE {\quad \textbf{Else} Generate two new nodes by setting an unfixed variable at $s_m$ and $s_{m+1}$}
\STATE {\quad\quad Go to a node that has not been visited or pruned}
\STATE {\quad \textbf{If} The current node has a corresponding SE $R_{curr}$:}
\STATE {\quad\quad \textbf{If} $R_{curr}>R_{lb}$: $R_{lb}=R_{curr}$}
\STATE {\textbf{Output:}$\bm{s}$;}
\end{algorithmic}
\end{algorithm}
\vspace{-.3cm}
\section{Simulation Results}\label{Simulation Sec}

In this section, we evaluate the performance of the IOS assisted system with the proposed algorithm in an indoor environment, and compare it with the IRS assisted system as proposed in~\cite{AZY2020,DZLSLH2020} and the conventional cellular system. In the IRS assisted system, the IRS only reflects the signals from the BS to the MU, and no transmissive signal is considered. In the conventional cellular system, we only consider the direct link from the BS to the MU. In our simulation, we set the height of the BS and the center of the IOS as 2 m, and the distance between the BS and the IOS being 500 m. The MU is randomly deployed within a circle of radius 2 m centering at the IRS, and the results present below are the average performance of over 10000 instances of the Monte Carlo simulation. The maximum transmit power of the BS $P_B$ is 40 dBm, the power of the AWGN is -96 dBm, and the IOS element separation is 0.03 m. The power ratio of the transmissive signal to the reflective signal $\epsilon$ is set as 1, the number of phase shifts for each IOS element is set as $S_a=4$.

Fig.~\ref{simulation0} shows the maximum SE of the MU on different locations, with the BS on the left side of the surface. An IOS/IRS with the length of 4 m and the height of 0.6 m is deployed at the line segment ((0,-2),(0,2)), and the BS is deployed at (-500,0). In the IOS system, the SE on either side of the surface can be improved. A higher SE can be obtained when the MU is closer to the center of the IOS, where the reflective-transmissive channel has a better quality. In the IRS system, the MU has a high SE only when it is on the left side of the surface.

\begin{figure}[!t]
    \centering
    \includegraphics[width=2.6in]{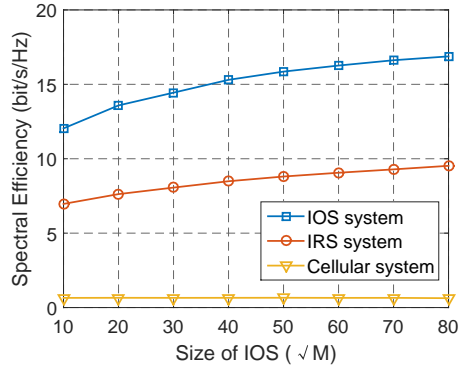}\vspace{-3mm}
    \caption{Size of IOS vs. SE.}\vspace{-7mm}
    \label{simulation1}
\end{figure}

Fig.~\ref{simulation1} depicts the relation between the average SE of the MU and the size of the IOS. The average SE is defined as the expectation of SE $R$ with the random distribution of the MU. The IOS is considered as a square array with $\sqrt{M}$ elements on each line and each row. The SE increases with the number of IOS elements, and the growth rate gradually slows down with the IOS size. An IOS with $10\times10$ elements improves the average SE for about 20 times when compared to the conventional cellular system, while an IRS of the same size only improves the average SE for about 12 times. The performance difference between the two systems is caused by the service coverage. The IOS can improve the average SE of the MU on either side of the surface, while the IRS can only improve that of the MU on one side.

%\begin{figure}[t]
%    \centering
%    \includegraphics[width=3in]{single-Sa-rate.eps}\vspace{-3mm}
%    \caption{Number of available states for each IOS element~($S_a$) vs. Sum-rate.}\vspace{-3mm}
%    \label{simulation2}
%\end{figure}

%Fig.~\ref{simulation2} shows the change of sum-rate with respect to the number of available states for each IOS element $S_a$. The sum-rate increases with $S_a$ and converges to a stable value with a large number of available states for each IOS element. The performance gain increases for about 20\% when $S_a$ changes from 2 to 4 due to more freedom of IOS configuration, but only increases for about 5\% when $S_a$ changes from 4 to 16. Since the complexity grows exponentially with $S_a$, it is not necessary to set $S_a$ as larger as possible. The growth of sum-rate of IOS is larger than IRS, since the IOS serves more MUs than the IRS, which benefits more from the freedom of IOS configuration.
\vspace{-3mm}
\section{Conclusions}\label{Conclusion Sec}
In this paper, we have studied an IOS-assisted downlink system. The IOS is capable to enhance the received signal of the MU on either side of it with IOS phase shift design. We have formulated an IOS phase shift design problem to maximize the SE of the system. The optimal phase shifts of the IOS elements have been solved, and an algorithm that designs the IOS phase shift in the finite set has been proposed. Simulation results have shown that the IOS significantly extends the service coverage of the BS when compared to the IRS. An IOS with a larger size can provide a higher SE for the MU, while the growth rate reduces with more IOS elements.
\vspace{-5mm}
\begin{appendices}
\section{Proof of Proposition 1}
Given the location of the BS, IOS, and MU, the SE maximization problem can be simplified as maximizing the channel gain, and the relaxed problem can be expressed as
\begin{equation}\label{appendix equation}
\begin{split}
\max_{\bm{s}} & \ \mathbb{E}\left(|h|^2\right),\\
s.t. \ & 0\leq\psi_m<2\pi, m=1, 2, \cdots M.
\end{split}
\end{equation}

We then substitute~(\ref{direct path equation}) and~(\ref{BS-IOS-MU channel}) into~(\ref{channel equation}). Given that $h_m^{NLoS}$ is zero mean and is independent for different elements, (\ref{appendix equation}) can be converted to
$\mathbb{E}\left(|h|^2\right)=\sum_{m=1}^M \left(\frac{\kappa}{1+\kappa}A\exp\Big(\frac{-j2\pi(d_{BS,m}+d_{m,MU})}{\lambda}\Big)\exp{(-j \psi_{m})}+\frac{1}{1+\kappa}\right.$ $\left.PL(d_{BS,m})PL(d_{m,MU})\right)
+\frac{\kappa}{1+\kappa}PL^{LoS}\exp\Big(\frac{-j2\pi(d_{BS,MU})}{\lambda}\Big)+\frac{1}{1+\kappa}PL(d_{BS,MU}),$
where $A=\frac{\lambda \sqrt{G^{tx} F_{m}^{tx} G^{rx} F^{rx}}\times \sqrt{G_m K^{A}(m) K^{D}(m)\delta_x\delta_y|\gamma_m|^2} }{(4\pi)^{\frac{3}{2}}d_{BS,m}d_{m,MU}}$ is not affected by the phase shift of the IOS. $PL^{LoS}$ is the LoS pathloss of the direct link, and $PL(d_{BS,MU})$ is the NLoS pathloss of the direct link, which are also not affected by the phase shift of the IOS. To maximize~(\ref{appendix equation}), the phase shift term of each element $\exp\Big(\frac{-j(2\pi(d_{BS,m}+d_{m,MU})+\psi_{m})}{\lambda}\Big)$ should be consistent with the direct path, i.e., $\psi_m+\frac{2\pi(d_{BS,m}+d_{m,MU})}{\lambda}=\frac{2\pi(d_{BS,MU})}{\lambda}$. Therefore, the optimal solution of element $m$ satisfies
\vspace{-1mm}
\begin{equation}\vspace{-1mm}
\psi_m=\frac{2\pi}{\lambda}(d_{BS,MU}-d_{BS,m}-d_{m,MU}).
\end{equation}

\end{appendices}
%\vspace{-2mm}


\begin{thebibliography}{11}\vspace{-1mm}
%\bibitem{SYPBK2017}
%D. Smith, O. Yurduseven, L. Pulido-Mancera, P. Bowen, and N. Kundtz. ``Analysis of a waveguide-fed metasurface antenna," \emph{Phys. Rev. Applied,} vol. 8, no. 5, pp. 1-16, Nov. 2017.
\bibitem{R2019}
M. Renzo, M. Debbah, D. Phan-Huy, A. Zappone, M. Alouini, C. Yuen, V. Sciancalepore, G. C. Alexandropoulos, J. Hoydis, H. Gacanin, J. Rosny, A. Bounceur, G. Lerosey, and M. Fink, ``Smart Radio Environments Empowered by Reconfigurable AI Meta-surfaces: An Idea Whose Time Has Come," \emph{EURASIP J. Wireless Commun. Netw.}, vol. 2019, no. 129, pp. 1-20, May 2019.
%\bibitem{SDEYIS2019}
%N. Shlezinger, O. Dicker, Y. Eldar, I. Yoo, M. Imani, and D. Smith, ``Dynamic metasurface antennas for Uplink massive MIMO systems", \emph{IEEE Trans. Commun.,} vol. 67 no. 10, pp. 6829-6843, Oct. 2019.
\bibitem{EZSSHL2020}
M. A. Elmossallamy, H. Zhang, L. Song, K. Seddik, Z. Han, and G.~Y.~Li, ``Reconfigurable intelligent surfaces for wireless communications: Principles, challenges, and opportunities," \emph{IEEE Trans. Cogn. Commun. Netw.,} vol. 6, no. 3, pp. 990-1002, Sep. 2020.
\bibitem{HZDLSLHP2019}
J. Hu, H. Zhang, B. Di, L. Li, L. Song, Y. Li, Z. Han, H. V.~Poor, ``Reconfigurable intelligent surfaces based radio-frequency sensing: Design, optimization, and implementation," \emph{IEEE J. Sel. Areas Commun., (Early Access),} Jul. 2020.
\bibitem{HZADY2019}
C. Huang, A. Zappone, G. C. Alexandropoulos, M. Debbah, and C. Yuen, ``Reconfigurable intelligent surfaces for energy efficiency in wireless
communication," \emph{IEEE Trans. Wireless Commun.,} vol. 18, no. 8, pp. 4157-4170, Aug. 2019.
\bibitem{ZDSH2020}
H. Zhang, B. Di, L. Song, and Z. Han, ``Reconfigurable intelligent surfaces assisted communications with limited phase shifts: how many phase shifts are enough?" \emph{IEEE Trans. Veh. Technol.,} vol. 69, no. 4, pp. 4498-4502, Apr. 2020.

%\bibitem{CYBF2017}
%Q. Chen, S. Yang, J. Bai, and Y. Fu, ``Design of absorptive/transmissive frequency-selective surface based on parallel resonance," \emph{IEEE Trans. Antennas Propag.,} vol. 65, no. 9, pp. 4897-4902, Sep. 2017.

\bibitem{AB2019}
V. Arun and H. Balakrishnan, ``RFocus: practical beamforming for small devices," [Online] Available: https://arxiv.org/abs/1905.05130
\bibitem{TCCDHRZJCC2019}
W. Tang, M. Chen, X. Chen, J. Dai, Y. Han, M. Renzo, Y. Zeng, S. Jin, Q. Cheng, and T. Cui ``Wireless communications with reconfigurable intelligent surface: path loss modeling and experimental measurement," [Online] Available: https://arxiv.org/abs/1911.05326
\bibitem{PG2013}
C. Pfeiffer, and A. Grbic, ``Metamaterial Huygens' surfaces: tailoring wave fronts with reflectionless sheets," \emph{Phys. Rev. Lett.,} vol.~110, no.~197401, pp.~1-5, May 2013.
\bibitem{G2005}
A. Goldsmith, \emph{Wireless Communications,} Cambridge Press, 2005.

\bibitem{DZSLHP2019}
B. Di, H. Zhang, L. Song, Y. Li, Z. Han, and H. V. Poor, ``Hybrid beamforming for reconfigurable intelligent
surface based multi-user communications: achievable rates with limited discrete phase shifts," \emph{IEEE J. Sel. Areas Commun.,} vol. 38, no. 8, pp. 1809-1822, Aug. 2020.
\bibitem{TV2005}
D. Tse and P. Viswanath, \emph{Fundamentals of Wireless Communications}, Cambridge Univ. Press, Cambridge, U.K., 2005.
\bibitem{AZY2020}
S. Abeywickrama, R. Zhang, and C. Yuen, ``Intelligent reflecting surface: Practical phase shift model and beamforming optimization." \emph{IEEE Trans. Commun., (Early Access),} Jun. 2020.
\bibitem{DZLSLH2020}
B. Di, H. Zhang, L. Li, L. Song, Y. Li, and Z. Han, ``Practical hybrid beamforming with finite-resolution phase shifters for reconfigurable intelligent surface based multi-user communications," \emph{IEEE Trans. Veh. Technol.,} vol. 69, no. 4, pp. 4565-4570, Apr. 2020.
\end{thebibliography}
\end{document}